\documentclass[prl,showpacs,twocolumn,amssymb]{revtex4}
\usepackage{latexsym}

\newcommand{\IR}{\mathbb{R}}

\newcommand{\epsi}{\varepsilon}
\begin{document}

\title{Space-adiabatic Decoupling to All Orders}

\author{Gianluca Panati}
\altaffiliation[Also at ]{Math.\ Phys.\ Sector, SISSA, Trieste}
\email{panati@sissa.it}

\author{Herbert Spohn}
 \email{spohn@ma.tum.de}

\author{Stefan Teufel}
 \email{teufel@ma.tum.de}

\affiliation{Zentrum Mathematik and Physik Department,
TU M\"unchen, D-80290 M\"unchen, Germany}


\date{January 28, 2002}

\begin{abstract}
 A systematic perturbation scheme is developed for
approximate solutions to the time-dependent Schr\"odinger equation
with a space-adiabatic Hamiltonian. For a particular isolated
energy band, the basic approach is to separate kinematics from
dynamics. The kinematics is defined through a subspace of the full
Hilbert space for which transitions to other band subspaces are
suppressed to all orders and the dynamics operates in that
subspace in terms of an effective intraband Hamiltonian. As novel
applications we discuss the Born-Oppenheimer theory to second
order and derive the nonperturbative definition of the $g$-factor
of the electron within nonrelativistic quantum electrodynamics.
\end{abstract}

\maketitle

The importance of slow parameter variations is inseparably linked
to early quantum mechanics \cite{Eh} and stood godfather for the
time-adiabatic theorem of Born and Fock \cite{BF}, later rigorized
by Kato \cite{Ka}. The physical idea is simple and best
demonstrated in the context of the dynamics of molecules. The
nuclei move slowly and follow, within a good approximation, some
classical trajectory. The electrons are then subject to a
potential of slow time variation. Abstracting from the particular
setting, one can think of a time-dependent Hamiltonian $H(t)$
acting on the Hilbert space $\mathcal H$. The slow time-dependence
is introduced through $H(\epsi t)$, where $\epsi \ll 1$ is a
dimensionless scale parameter, and one wants to study the solution
to the time-dependent Schr\"{o}dinger equation
\begin{equation}\label{a}
i \epsi \frac{d}{d t} \psi_t = H(t)\psi_t
\end{equation}
in the limit of small $\epsi$. For simplicity $\hbar = 1$, except
for the physical examples. In (\ref{a}) we followed the standard
practice to switch to the slow time scale through the substitution
$t/\epsi$ for $t$. It should only be recalled that $t$ of order
$1$ corresponds to very long times when measured in atomic units.
We assume that $H(t)$ has a physically distinguished ``relevant''
subspace of dimension $m$ with, only for the sake of presentation,
single energy $E(t)$, which is isolated from the rest of the
spectrum. The projection $P(t)$ onto the relevant subspace is
spanned by the instantaneous eigenvectors
$|\varphi_\alpha(t)\rangle$ of $H(t)$ satisfying
$H(t)|\varphi_\alpha(t)\rangle=E(t)|\varphi_\alpha(t)\rangle$,
$\alpha=1,...,m$. The initial wave function, $\psi_0$, is assumed
to lie in the relevant subspace, $\psi_0 =P(0)\psi_0$. There is no
reason to expect persistence as $\psi_t = P(t)\psi_t$ at later
times. The adiabatic theorem asserts however that such a property
holds approximately in the sense that
\begin{equation}\label{c}
\|(1 - P(t))\psi_t\| = {\mathcal O}(\epsi)
\end{equation}
with $\|\psi\| = \langle\psi|\psi\rangle^{1/2}$ the length of the
vector $\psi$.

While the error estimate (\ref{c}) is undoubtedly correct, it
really begs the question, since the nature of ${\mathcal
O}(\epsi)$ is left unspecified. It could be that a piece of size
$\epsi$ of $\psi_t$ leaks out into the orthogonal subspace $(1 -
P(t)){\mathcal H}$. Alternatively $|\psi_t\rangle\langle\psi_t|$
is slightly tilted relative to $P(t)$. As first recognized by
Lenard \cite{Le}, and on a more refined level in
\cite{AJS,JKP,Be,Ne}, it is the latter option which is realized by
the solution to (\ref{a}). There is then an iterative procedure
for constructing a projection $P^\epsi (t)$, $\epsi$-close to
$P(t)$, such that $\psi_t$ remains in $P^\epsi (t){\mathcal H}$ up
to any order, in symbols
\begin{equation}\label{d}
\|(1 - P^\epsi(t))\psi_t\| = {\mathcal O} (\epsi^\infty)\,,\quad
\psi_0=P^\epsi(0)\psi_0\,.
\end{equation}
The power series for $P^\epsi(t)$ is asymptotic,
 with an error of order  $\exp[-1/\epsi]$ in case $H(t)$
is analytic in $t$ \cite{JP,JP1}.

Such results are of physical interest only if higher order
corrections can be computed in a systematic fashion. We explain a
scheme which naturally generalizes to the space-adiabatic
decoupling. The idea is to separate kinematics, the subspace
$P^\epsi(t){\mathcal H}$, from dynamics, the unitary evolution
inside $P^\epsi(t){\mathcal H}$ as generated by some effective
Hamiltonian. For this purpose we choose a reference subspace of
dimension $m$ with time-{\it independent} basis
$|\chi_\alpha\rangle\,,\alpha = 1,...,m.$ The effective
Hamiltonian operates in this subspace and to order $\epsi^2$ is
given by
\begin{eqnarray}\label{e}\lefteqn{
H^{\mathrm{eff}}_{\alpha\beta}(t)  =   \delta_{\alpha\beta} E(t)
-i \epsi  \langle\varphi_\alpha (t)|\dot\varphi_\beta (t)\rangle}
\\&&\hspace{-0.3cm}
 + \frac{1}{2} \epsi^2 \langle\dot\varphi_\alpha (t)|(H(t) -
E(t))^{-1} (1-P(t))\dot\varphi_\beta (t)\rangle
+ {\mathcal O}(\epsi^3)\,, \nonumber
\end{eqnarray}
$\alpha,\beta = 1,...,m$. Note that the dynamical problem
(\ref{a}) has been reduced to the dimension of the relevant
subspace. Secondly there is a unitary, $U^\epsi (t),$ which
rotates the reference subspace into $P^\epsi (t){\mathcal H}.$ To
order $\epsi$ one finds
\begin{eqnarray}\label{f}\lefteqn{
U^\epsi (t) = \sum\limits_{\alpha = 1}^m \Big(|\varphi_\alpha
(t)\rangle}\\&& + i \epsi (H (t) - E (t))^{-1} (1 - P
(t))|\dot\varphi_\alpha (t)\rangle\Big) \langle\chi_\alpha| +
{\mathcal O}(\epsi^2)\,.\nonumber
\end{eqnarray}
Thus, if $\chi^\epsi (t)$ denotes the time-evolved vector in the
reference subspace, i.e.\ the solution to $i \epsi d \chi^\epsi
(t)/d t = H^{\mathrm {eff}} (t)
\chi^\epsi(t)\,,\chi^\epsi(0)=\chi$, then the $\mathcal
O(\epsi^\infty)$-approximate solution to (\ref{a}) reads
\begin{equation}\label{g}
\psi_t = U^\epsi (t) \chi^\epsi (t)\,, \quad\psi_0 = U^\epsi (0)
\chi\,,
\end{equation}
which is valid on the slow time scale with an error ${\mathcal
O}(\epsi^2)$ upon inserting (\ref{e}) and (\ref{f}).

Physically, the separation into prescribed slow degrees of freedom
and rapidly adjusting fast degrees  of freedom is somewhat
artificial. In actual fact there is a single time-independent
Hamiltonian governing all degrees of freedom. For example, the
molecular Hamiltonian is independent of time and the separation
into slow nuclei and fast electrons results from the small mass
ratio. What is then needed is a {\em spatial} version of the
time-adiabatic theorem in the sense that some dynamical degrees of
freedom are frozen out. 
While the physical mechanism behind such a space-adiabatic
decoupling may differ widely, our discovery is that there is a
unifying and general theoretical framework. We like to compare the
situation with standard bound state perturbation theory. It is
always the same algebra, but the physical manifestations are rich
and differ widely depending on the context. The general scheme to
be sketched below is applicable to a variety of systems such as
the Dirac equation with slowly varying external fields, the
Born-Oppenheimer Hamiltonian, electron motion in periodic
potentials, magnetic Bloch bands, motion on embedded manifolds
with constraining transverse potential, and others. These systems
have been studied before by model dependent methods and it is not
uncommon to have a debate as regards to the structure of second
order corrections \cite{LW, AS, CN, CN1, Ba}. In contrast our
perturbation scheme systematically computes higher order
corrections and provides error bounds. Such a scheme was
envisioned by E.\ I.\ Blount in a remarkable series of papers
\cite{Bla,Blb,Blc}. Particular aspects are studied by Littlejohn
and Flynn \cite{LF, LF1} and Nenciu and Sordoni \cite{NS}. The
full justification of the space-adiabatic perturbation theory is
beyond the size of a letter and given elsewhere \cite{PST}. Here
we merely explain the structure of the leading order terms. As
novel applications we discuss the second order Born-Oppenheimer
theory and derive the gyromagnetic ratio ($g$-factor) for an
electron coupled to the quantized radiation field within
nonrelativistic quantum electrodynamics.

We claim that the spatial version of Eq.\ (\ref{a}) reads
\begin{equation}\label{h}
i \epsi\frac{\partial}{\partial t} \psi_t = H ({\widehat q}, \epsi
{\widehat p}\,) \psi_t\,.
\end{equation}
$q\in {\IR }^d$ is a position like variable and
$p$ the corresponding canonical momentum. 
Again, $\epsi\ll 1$ is a dimensionless scale  parameter, which controls
the variation of $q$ and $p$.
The phase space for
the slow degrees of freedom is $\Gamma = {\IR}^{2d}.$ $H (q,p)$ is
an operator valued function on $\Gamma$, which acts on the Hilbert
space ${\mathcal H}_{\mathrm f}$ of ``internal" degrees of
freedom, and ${\widehat q} = x,{\widehat p} = -i\nabla_x.$ To
properly define $H ({\widehat q},\epsi {\widehat p}\,)$ we adopt
the Weyl rule for operator ordering, 
\begin{eqnarray}\label{i}\hspace{-1cm}
H ({\widehat q},\epsi {\widehat p}\,) \psi(x) &=& (2\pi)^{-d} \int
d^d x' d^d\xi\, e^{i(x-x')\cdot \xi}\nonumber\\
&&\qquad\times\, H({\textstyle\frac{1}{2}}
(x+x'), \epsi\xi) \psi(x')\,,
\end{eqnarray}
where $H$ acts as an operator on $\psi$. The ``spinor'' wave
function $\psi(x)$ takes values in ${\mathcal H}_{\mathrm f}$.
Thus the Hilbert space for the Schr\"odinger equation (\ref{h}) is
${\mathcal H} = L^2 ({\IR}^d,{\mathcal H}_{\mathrm f}) =
L^2({\IR}^d)\otimes{\mathcal H}_{\mathrm f}.$

The eigenvalues of the Hamiltonian $H(q,p)$ define the energy
bands. One of them, or one group of them, is considered to be
physically relevant and, only for the sake of simplified
presentation, is assumed to consist of a single band energy $E(q,p)$ of
constant multiplicity $m$. Thus there are $m$ orthogonal
eigenvectors satisfying $H (q,p)|\psi_\alpha(q,p)\rangle = E
(q,p)|\psi_\alpha(q,p)\rangle$. They span the projector
$P(q,p)=\sum_{\alpha=1}^m |\psi_\alpha
(q,p)\rangle\langle\psi_\alpha (q,p)|.$ As crucial assumption one
needs that for each $q,p$ the energy $E(q,p)$ is separated by a
gap from the remainder of the spectrum of $H(q,p)$. In spirit, to
lowest order in $\epsi$, the projector onto the adiabatically
decoupled subspace of ${\mathcal H}$ is the Weyl quantization
$P({\widehat q},\epsi{\widehat p}\,)$. However, $P({\widehat
q},\epsi{\widehat p}\,)$ is not a projector, in general. Still on
abstract grounds one can construct a time-independent subspace
$P^\epsi{\mathcal H}$, the projection $P^\epsi$ being 
$\epsi$-close to $P({\widehat q},\epsi{\widehat p}\,)$, which is
adiabatically decoupled to all orders in $\epsi$, in the sense
that
\begin{equation}\label{j}
\|(1-P^\epsi)\psi_t \| = {\mathcal O}(\epsi^\infty)\,,\quad \psi_0
= P^\epsi\psi_0\,,
\end{equation}
for $\psi_t$ the solution to (\ref{h}), compare with (\ref{d}).

To systematically compute approximate solutions we disentangle
kinematics and dynamics. The reference Hilbert space is now
$L^2{(\IR}^d)\otimes {\mathbb C}^m$ with fixed spinor basis
$|\chi_\alpha\rangle,\alpha=1,...,m$. The effective Hamiltonian to
order $\epsi$ turns out to be the Weyl quantization of
\begin{eqnarray}\lefteqn{\label{k}
H^{\mathrm{eff}}_{\alpha\beta} (q,p) = \delta_{\alpha\beta}
E(q,p)} \\&& -\textstyle\frac{1}{2}i\, \epsi
\langle\psi_\alpha (q,p)|\{H(q,p) + E(q,p),\psi_\beta(q,p)\}
\rangle + {\mathcal O}(\epsi^2)\,.\nonumber
\end{eqnarray}
$\{\cdot, \cdot \}$ is  the  Poisson bracket defined for the operator-
resp.\ vector-valued functions $A$, $\psi$
as $\{A,\psi\}=\nabla_ p A\cdot\nabla_q\psi-
\nabla_ q A\cdot\nabla_ p \psi$. The reference subspace is rotated 
by the unitary
$U^\epsi$ into the adiabatically decoupled subspace
$P^\epsi{\mathcal H}.$ To lowest order one obtains the Weyl
quantization of
\begin{equation}\label{l}
U_0(q,p) = \sum_{\alpha=1}^m
|\psi_\alpha(q,p)\rangle\langle\chi_\alpha|\,.
\end{equation}
$U_0({\widehat q},\epsi{\widehat p}\,)$ is not unitary in general,
but the error is in higher orders. As in the time-adiabatic
setting the approximate solution to the Schr\"odinger equation
(\ref{h}) is given by
\begin{equation}\label{la}
\psi_t=U^\epsi(\widehat q,\epsi\widehat p\,)\exp[-i
H^{\mathrm{eff}}(\widehat q,\epsi\widehat
p\,)t/\epsi]\chi_0+\mathcal{O}(\epsi^\infty)\,,
\end{equation}
where $\psi_0=U^\epsi(\widehat q,\epsi\widehat p\,)\chi_0$.
Inserting from (\ref{k}) and (\ref{l}), the solution holds on the slow
time scale with an error ${\mathcal O}(\epsi)$.

Higher order corrections to (\ref{k}), (\ref{l}) are available
\cite{PST}, but they rapidly increase in complexity. To provide
some idea we explain the second order Born-Oppenheimer
approximation, for which the Hamiltonian has the form
\begin{equation}\label{m}
H_{\mathrm{BO}}(q,p)=\textstyle\frac{1}{2}p^2\mathbf{1}+V(q)\,,
\end{equation}
where $\mathbf{1}$ is the unit operator and $V(q)$ is some
operator both acting on $\mathcal H_{\mathrm f}$. For a molecule,
$q$ would stand for the coordinates of the nuclei and $V(q)$ is
the electronic Hamiltonian with the nuclei fixed at $q$. The
adiabaticity parameter is
$\epsi=(m_{\mathrm{el}}/m_{\mathrm{nuc}})^{1/2}$, assuming that
the nuclei have all the same mass, and $H_{\mathrm{BO}}(\widehat
q,\epsi\widehat p\,)=-\epsi^2\Delta/2+V(x)$ is the standard
molecular Hamiltonian. The relevant subspace is spanned by
eigenvectors to $V(q)$,
$V(q)|\psi_\alpha(q)\rangle=e(q)|\psi_\alpha(q)\rangle\,,
\alpha=1,...,m$. Thus $E(q,p)=\textstyle\frac{1}{2}p^2+e(q)$ and
$P(q,p)$ depends only on $q$ as
$P(q)=\sum_{\alpha=1}^m|\psi_\alpha(q)\rangle\langle
\psi_\alpha(q)|$, which implies that also $P(\widehat q\,)$ is a
projector without additional corrections from the
operator ordering. Generalizing \cite{LW,WL}, the effective
Hamiltonian including second order is computed to
\begin{eqnarray}\label{n} \lefteqn{
H_{\mathrm{BO}\alpha\beta}^{\mathrm{eff}}(q,p)=\textstyle
\frac{1}{2}\sum
\limits_{\gamma=1}^{m}
(p\delta_{\alpha\gamma}-\epsi
A_{\alpha\gamma}(q))\cdot(p\delta_{\gamma\beta}-\epsi
A_{\gamma\beta}(q))}\nonumber\\
&& +\,e(q)\delta_{\alpha\beta}
+\textstyle\frac{1}{2}\epsi^2\langle\nabla\psi_\alpha(q)|
\cdot\nabla
\psi_\beta(q)\rangle \nonumber\\
&&-\,\epsi^2\langle
p\cdot\nabla\psi_\alpha(q)|(V(q)-e(q))^{-1}(1-P(q))p\cdot
\nabla\psi_\beta(q)
\rangle \nonumber\\&&
+\,\mathcal O(\epsi^3)\,,
\end{eqnarray}
where the ``vector potential" of the Berry connection reads
$A_{\alpha\beta}(q)=i\langle\psi_\alpha(q)|\nabla\psi_\beta(q)
\rangle$.
As before, the effective quantum Hamiltonian to that order is the
Weyl quantization $H^{\mathrm{eff}}_{\mathrm{BO}}(\widehat
q,\epsi\widehat p\,)$. For the unitary, which rotates the
reference space back to $P^\epsi{\mathcal H}$, one obtains
\begin{eqnarray}\label{nb}\lefteqn{\hspace{-1cm}
U_{\mathrm{BO}}(q,p)=\sum\limits_{\alpha=1}^m\Big(|
\psi_\alpha(q)\rangle+i \epsi\big(V(q)-e(q)\big)^{-1}}
\nonumber\\&&\times\,
(1-P(q))|p\cdot\nabla\psi_\alpha(q)\rangle\Big)\langle\chi_\alpha|
+{\mathcal O}(\epsi^2)\,.
\end{eqnarray}
Note the similarity with the time-adiabatic approximation
(\ref{e}), (\ref{f}).

We skipped over the issue on which time scale (\ref{la}) is valid.
Clearly, there are two different ways of reading our result.
Either one considers the time scale $\mathcal O(\epsi^0)$ on which
(\ref{h}) is written. Then, inserting (\ref{n}) and (\ref{nb}),
 the error in (\ref{la}) is $\mathcal
O(\epsi^2)$. Or one considers the longer time scale $\mathcal
O(\epsi^{-1})$. Then precision is lost and the error is guaranteed
to be $\mathcal O(\epsi)$ only. Note that in this second variant
the adiabatic decoupling time is much longer than the one which
would result from a semiclassical analysis of the Hamiltonian
$H_{\mathrm{BO}}^{\mathrm{eff}}(\widehat q,\epsi\widehat p)$. Its
validity is restricted to the Ehrenfest time which usually is of
order $\log\epsi^{-1}$ in our units.

To have a concrete physical example we discuss the gyromagnetic
ratio of the electron in nonrelativistic QED. The electron is
subject to a weak uniform external magnetic field. The internal
degrees of freedom are the photons which rapidly adjust to a
state of lowest energy consistent with the momentary state of the
electron. The orbital motion is circular with frequency $\omega_c$
and the spin motion is precession with frequency $\omega_s$. As
the standard definition of the $g$-factor we adopt
\begin{equation}\label{nc}
g = 2\omega_s/\omega_c\,.
\end{equation}

To apply our general scheme it is convenient to consider first the
motion in arbitrary external potentials of slow variation. The
$g$-factor will be retrieved as a special case. In dimensionless
Heaviside--Lorentz units the Hamiltonian with zero external
potentials reads
\begin{equation}\label{o}
H = (1/2) \big(\sigma\cdot (p_{\mathrm{el}} - e A_\varphi
(x))\big)^2 + H_{\mathrm f}\,,
\end{equation}
$\alpha=e^2/4\pi$ the fine-structure constant \cite{Mi}.
$x,p_{\mathrm{el}}$ are position and momentum of the electron,
$\sigma$ is the vector of the Pauli matrices,
$H_{\mathrm f} = \sum_{\lambda = 1}^2 \int d^3 k |k|
a^{\dagger} (k,\lambda) a (k,\lambda)$ is the energy of the
photons with dispersion $\omega (k) = |k|$ and helicity $\lambda$,
and $A_\varphi(x)$ is the quantized transverse vector potential,
which is smeared with the form factor $\varphi$ (=ultraviolet
cutoff). The dressed electron responds to the slowly varying
external potentials $\phi_{\mathrm{ex}} (\epsi x), A_{\mathrm{ex}}
(\epsi x)$ through a slow variation of the total momentum $p =
p_{\mathrm{el}}+P_{\mathrm f}$, where $P_{\mathrm f} =
\sum_{\lambda = 1,2} \int d^3 k k a^\dagger (k,\lambda) a
(k,\lambda)$ is the momentum of the photons. To write the
Hamiltonian in the standard space-adiabatic form (\ref{h}) one
thus has to substitute $p - P_{\mathrm f}$ for $p_{\mathrm{el}}$
and $x/\epsi$ for $x$ in order to arrive at the operator valued
function
\begin{eqnarray}\label{p}
H (q,p) &=& {\textstyle\frac{1}{2}} \big(p - P_{\mathrm f} - e
A_\varphi - e A_{\mathrm{ex}} (q)\big)^2 \\&&-\, {\textstyle
\frac{1}{2}}
e \big(B_\varphi + \epsi B_{\mathrm{ex}} (q)\big) \cdot \sigma +
e\phi_{\mathrm{ex}} (q) + H_{\mathrm f}\nonumber
\end{eqnarray}
with the abbreviations $A_\varphi=A_\varphi (0),\,
B_\varphi=\nabla\times A_\varphi (0),B_{\mathrm{ex}} (q) =
\nabla\times A_{\mathrm{ex}}(q)$. The Weyl quantization of
(\ref{p}) is the Hamiltonian of nonrelativistic QED. $H(q,p)$ acts
on the electron spinor and the wave function for the photons. Thus
${\mathcal H}_{\mathrm f} = {\mathbb C}^2 \otimes {\mathcal F}$
with ${\mathcal F}$ the photon Fock space. Note that $H (q,p)$ has
a rather particular structure which we abbreviate as
\begin{equation}\label{q}
H (q,p)=D (\widetilde {p}) - {\textstyle\frac{1}{2}} e \epsi
B_{\mathrm{ex}} (q) \cdot\sigma + e\phi_{\mathrm{ex}}(q)\,,
\end{equation}
where $\widetilde {p}=p-e A_{\mathrm{ex}}(q)$ is the kinetic momentum.

The physically relevant subspace is spanned by the eigenvectors of
lowest energy. They satisfy $D(\widetilde{p})|\psi_\alpha
(\widetilde {p})\rangle =
E_0(\widetilde{p})|\psi_\alpha(\widetilde{p})\rangle$, which
implies
$E(q,p)=E_0(p-eA_{\mathrm{ex}}(q))+e\phi_{\mathrm{ex}}(q)$. By
spin degeneracy the ground state subspace turns out to have
dimension $2$, i.e. $m=2,\alpha=\pm 1$ \cite{HS}. In addition,
$E_0(p)$ is isolated from the rest of the spectrum for $|p|$
smaller than Compton momentum, provided the interaction is cut off
in the infrared. We want to fix a basis such that $\psi_+$
corresponds to the electron spin pointing along the $+ z$
direction. For this purpose one notes that the total angular
momentum $J$ in the direction of $p$ is conserved, $[D(p),p\cdot
J] = 0$. Thus the eigenvectors along $\pm p$ are uniquely fixed
and by suitable linear combination therefore also along $\pm z$.

At this stage one merely inserts in (\ref{k}). The effective
Hamiltonian including first order is computed to
\begin{eqnarray}\label{r}\lefteqn{
H^{\mathrm{eff}}_{\alpha\beta}(q,p)=\Big(E_0(p - e
A_{\mathrm{ex}}(q)) + e \phi_{\mathrm{ex}}(q)\Big)
\delta_{\alpha\beta}}\nonumber\\
&&-\textstyle\frac{1}{2}\epsi e B_{\mathrm{ex}}
(q)\cdot\langle\psi_\alpha(\widetilde
p)|\sigma\psi_\beta(\widetilde
p)\rangle\nonumber\\
&&-\epsi e\big(- \nabla\phi_{\mathrm{ex}}(q) +  v \times
B_{\mathrm{ex}}(q)\big)\cdot\langle\psi_\alpha(\widetilde
p)|i\nabla\psi_\beta(\widetilde p)\rangle\nonumber\\
&&-{\textstyle\frac{1}{2}} i\epsi e\,
B_{\mathrm{ex}}(q)\cdot\langle\nabla\psi_\alpha(\widetilde
p)|\times (D(\widetilde p)-E_0(\widetilde
p))\nabla\psi_\beta(\widetilde p)\rangle
\nonumber\\&&+\,\mathcal O(\epsi^2)
\nonumber\\
&=&H^{\mathrm{eff}}_{0\alpha\beta}(q,p)+\epsi
H^{\mathrm{eff}}_{1\alpha\beta}(q,p)+\mathcal O(\epsi^2)\,,
\end{eqnarray}
where $\langle\cdot|\cdot\rangle$ is the scalar product in
${\mathbb C}^2\otimes{\mathcal F}$, ${\widetilde p} = p - e
A_{\mathrm {ex}}(q)$ as before, $v = \nabla E _0{(\widetilde p)}.$
Upon Weyl quantization $H^{\mathrm{eff}}_0$ is the Peierls
substitution. $H^{\mathrm{eff}}_{1}$ describes the spin
precession. The effective Hamiltonian (\ref{r}) is most
conveniently studied through semiclassical methods \cite{PST}. The
center $q_t,p_t$ of the wave packet is governed by the classical
Hamiltonian $E_0(p-e A_{\mathrm{ex}}(q)) +
e\phi_{\mathrm{ex}}(q).$ With our choice of time scale, $q_t,p_t$
no longer depend on $\epsi$. Inserting a particular orbit
$q_t,p_t$ in $H_1^{\mathrm{eff}}$ and cancelling the
$\epsi$-factors in (\ref{h}) and (\ref{r}) yields the spin
precession as
\begin{equation}\label{s}
i\frac{d}{dt}\chi_t=H_1^{\mathrm{eff}}(q_t, p_t)\chi_t\,.
\end{equation}
Orbital motion and spin precession vary on the same time scale and
the spin does not react back on the orbit to our order.

Of particular interest is a uniform external magnetic field,
$B_{\mathrm{ex}}(q) = B,\phi_{\mathrm{ex}}(q)=0,$ and small
velocities $v=\nabla E_0(\widetilde p)$, which means $\widetilde
p_t=0$ in good approximation. Let us introduce the bare mass $m$
of the electron, the velocity of light $c$, and Planck's constant
$\hbar$. Then the spin Hamiltonian becomes
\begin{eqnarray}\label{t}\lefteqn{
H^{\mathrm{spin}}_{\alpha\beta} = {-\frac{e\hbar}{2m}}
B\cdot\Big(\langle\psi_\alpha(0)|\sigma\psi_\beta(0)\rangle}
\\&&\hspace{-3mm}+\,i
\langle\nabla
\psi_\alpha(0)|
\times(D(0)-E(0))\nabla\psi_\beta(0)\rangle\Big)
=\textstyle\frac{1}{2}\hbar\mu
\left(B\cdot\sigma\right)_{\alpha\beta} \nonumber
\end{eqnarray}
which implicitly defines the frequency $\omega_s=\mu|B|$ of the
spin precession. On the other hand the orbital motion is
\begin{equation}\label{u}
m_{\mathrm {eff}} \ddot{q} = \frac{e}{c} \dot{q}\times B\,,\quad
m_{\mathrm {eff}}=1/E_0''(0)\,,
\end{equation}
which yields the cyclotron frequency $\omega_c=e|B|/m_{\mathrm
{eff}}c.$ Together with the definition (\ref{nc}) one deduces the
nonperturbative expression of the $g$-factor valid for any value
of the fine structure constant.

The matrix elements in (\ref{t}) and for $E_0''(0)$ are not
available as closed formulae. The only possibility is to expand in
$e$ around $e=0$. To second order in $e$ and removing the
ultraviolet cutoff one obtains
\begin{equation}\label{v}
g = 2 \big(1+ \frac{8}{3}(\frac{\alpha}{2\pi})+ {\mathcal
O}(\alpha^2)\big)
\end{equation}
in agreement with \cite{GK}, who use static second order
perturbation for the Zeeman splitting.

In summary, the expressions (\ref{k}) and (\ref{l}) and their
higher order corrections provide a systematic approximation in
space-adiabatic situations. To compute them one needs as explicit
as possible the eigenbasis in the relevant subspace. Just like in
static perturbation theory for eigenvalues at second order, also here
the inverse operator $(H(q,p)-E(q,p))^{-1}$ appears, compare with
(\ref{n}), (\ref{nb}), which is not necessarily readily available.
From our own experience approximations beyond second order become
tedious and most likely carry no additional physical insights.

\end{document}